\pdfoutput=1
\documentclass[sigconf]{acmart}
\usepackage{multirow}
\usepackage{tabularx,ragged2e,booktabs,caption}

\usepackage{mathrsfs}%
\usepackage{subcaption}
\usepackage{amsthm}
\usepackage{graphicx}
\usepackage{multicol}
\usepackage{placeins}

\usepackage{float}
\usepackage{dblfloatfix}
\usepackage{longtable}
\usepackage{supertabular}
\usepackage{amsmath}

\usepackage{booktabs} 
\usepackage{dsfont}

\newtheorem*{assumption*}{\assumptionnumber}
\providecommand{\assumptionnumber}{}
\makeatletter

\usepackage{times}
\usepackage{soul}
\usepackage{url}
\usepackage[utf8]{inputenc}
\usepackage{caption}
\usepackage{amsmath}
\usepackage{booktabs}
\usepackage{algorithm}
\usepackage{algorithmic}
  \usepackage{mathrsfs}
\usepackage{scalerel,amssymb}

\usepackage{braket}
\usepackage[utf8]{inputenc}
\usepackage[english]{babel}
\usepackage{multirow}
\usepackage{multirow}
\usepackage{makecell}

\usepackage{xcolor}
\renewcommand\footnotetextcopyrightpermission[1]{}
\makeatother

\pdfoutput=1
\begin{document}
\title{PMD: A New User Distance for Recommender Systems}

\author{Yitong Meng}
\affiliation{%
  \institution{The Chinese University of Hong Kong}
  \streetaddress{Shatian, N.T., Hong Kong}
  \postcode{XXX}
}
\email{ytmeng@cse.cuhk.edu.hk}

\author{Weiwen Liu}
\affiliation{%
  \institution{The Chinese University of Hong Kong}
  \streetaddress{XXX}
  \postcode{XXX}
}
\email{wwliu@cse.cuhk.edu.hk}

\author{Benben Liao}
\affiliation{%
  \institution{Tencent Quantum Lab}
  \streetaddress{XXX}
  \postcode{XXX}
}
\email{bliao@tencent.com}

\author{Jun Guo}
\affiliation{%
  \institution{Tsinghua-Berkeley Shenzhen Institute}
  \streetaddress{XXX}
  \postcode{XXX}
}
\email{eeguojun@outlook.com  }

\author{Guangyong Chen}
\affiliation{%
  \institution{Tencent Quantum Lab}
  \streetaddress{XXX}
  \postcode{XXX}
}
\email{gycchen@tencent.com}




\begin{abstract}
Collaborative filtering, a widely-used recommendation technique, predicts a user's preference by aggregating the ratings from similar users. 
Traditional similarity measures utilize ratings of only co-rated items while computing similarity between a pair of users.
As a result, these measures cannot fully utilize the rating information and are not suitable for real world sparse data. To solve these issues, we propose a novel user distance measure named Preference Mover's Distance (PMD) which makes full use of all ratings made by each user. Our proposed PMD can properly measure the distance between a pair of users even if they have no co-rated items.
We show that this measure can
be cast as an instance of the Earth Mover’s Distance, a well-studied transportation problem for
which several highly efficient solvers have been
developed.
Experimental results show that PMD can help achieve superior recommendation accuracy than state-of-the-art methods, especially when training data is very sparse.
\end{abstract}

%
%

\setcopyright{none}
\acmConference[Conference]{}{XX}{XX, XXX}

\pdfoutput=1
\maketitle
\pdfoutput=1
\pdfoutput=1
\section{Introduction}

Collaborative filtering (CF) is one of the most widely-used user-centric recommendation techniques in practice~\cite{guo2013novel,zheng2010collaborative}. For a specific user, CF recommends items according to the preference of similar users. User similarity plays an important role in CF, including both memory-based~\cite{zheng2010collaborative}
and model-based~\cite{ma2011recommender} approaches. First, it serves as a criterion to select a group of most similar users whose ratings will form the basis of recommendations. Second, it is also used to weigh the users so that more similar users will have greater impact on recommendations. 

\begin{figure}[t]
 \centering
    \includegraphics[width=3.9cm]{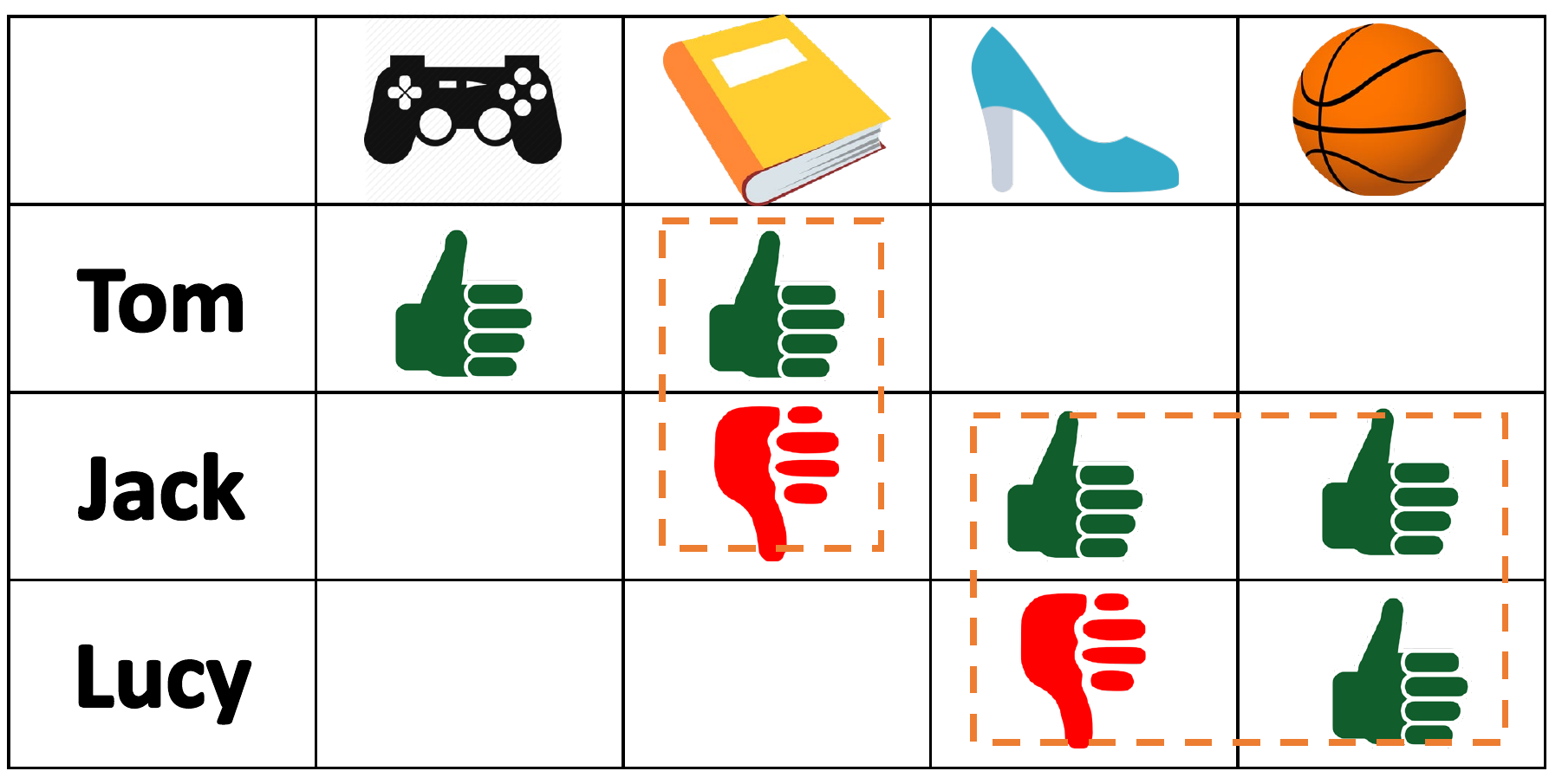}
    \vspace{-1.0em}
    \caption{Examples of co-rated items. Jack and Tom ```co-rated'' the book while Jack and Lucy ``co-rated'' the high-heeled shoes and the basketball. Tom and Lucy have no co-rated items.}
    \label{fig:co-rated}
      \vspace{-1.9em}
 \end{figure}

Some traditional similarity measures, such as Cosine (COS)~\cite{breese1998empirical}, Persons Correlation Coefficient (PCC)~\cite{breese1998empirical} and so on, have been widely used in CF to evaluate similarity~\cite{wang2017hybrid,desrosiers2011comprehensive}.
However, They only consider
ratings on the co-rated items~\cite{patra2015new,wang2017hybrid}, which may not represent the taste of a user properly. The reason is that the information is lost while ignoring the ratings on the non-co-rated items~\cite{patra2015new,wang2017hybrid}. Figure \ref{fig:co-rated} shows examples of co-rated items. Therefore, these measures perform poorly if there are no sufficient numbers of co-rated items and thus are not suitable for real world sparse data~\cite{patra2015new,wang2017hybrid}, because the more sparse the data the less likely the co-rated items can exist. Moreover, these measures are not applicable when there are no co-rated items at all.

To measure user similarity more accurately, one should make full use of all ratings of each user. However, computing user similarity based on the ratings of two different sets of items is challenging---let's imagine an extreme case that user A and B rate completely different items. 
Therefore, it requires additional information to build the connections between their ratings. 
We consider the similarity among items to establish the connection between ratings from different items.
Item similarity is actually more general than ``co-rated'', because ``co-rated'' means the users rate the same items and same is also one kind of similarity. 
We proposed to compute user similarity by assuming: if two users have similar opinions on similar items, then their tastes are similar. Our assumption differs from traditional methods~\cite{breese1998empirical} in that we compare user opinions based on similar items instead of just ``co-rated'' ones.

We propose the Preference Mover's Distance (PMD) which fully utilizes all ratings made by each user and can evaluate user similarity even in the absence of co-rated items. PMD can guarantee that if user A and B are both similar to C, then A and B should also be similar, as implied by triangle inequality.
The optimization problem of computing PMD reduces to a special case of the Earth Mover’s Distance~\cite{monge1781memoire,rubner1998metric,wolsey2014integer}, a well-studied transportation problem for which fast specialized solvers~\cite{ling2007efficient,pele2009fast} has been developed. Experimental results show that PMD can help achieve superior recommendation accuracy over state-of-the-art measures, especially when training data is sparse.
\pdfoutput=1

\begin{figure*}[t]
 \centering
    \includegraphics[width=11.1cm]{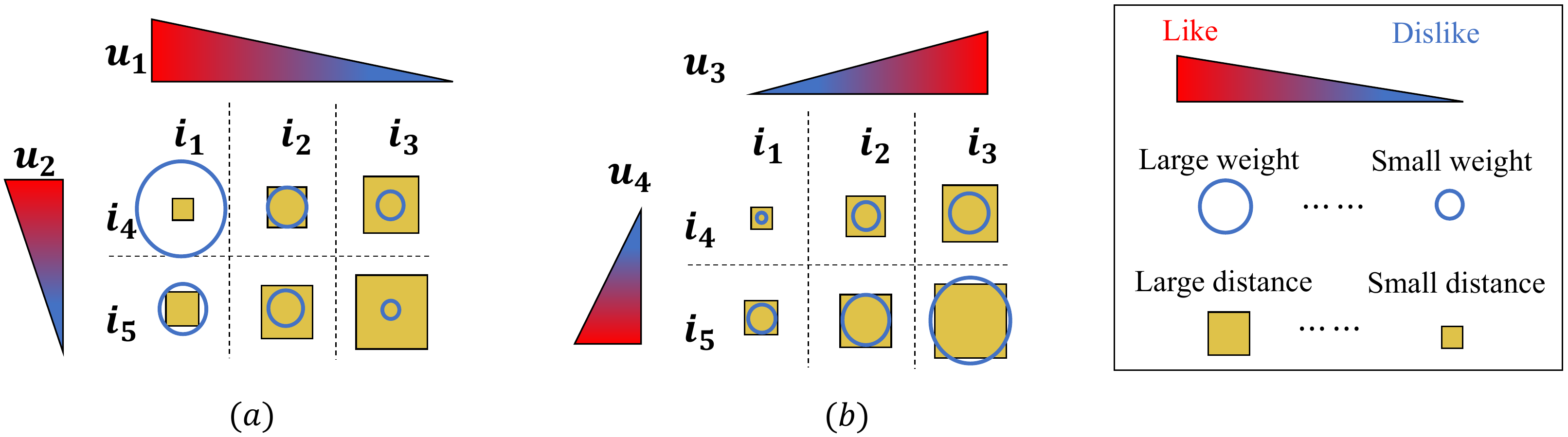}
    \caption{An illustration of our basic philosophy to evaluate user distance. 
    The bigger the square  the larger the item distance $d(i,j)$, e.g. $i_1$ and $i_4$ are similar while $i_3$ and $i_5$ are dissimilar. 
    The bigger the circle the larger the weight $\mathbf{W}_{u,v}(i,j)$. 
    In subfigure (a), $ \mathcal{I}_{u1}=\{i_1,i_2,i_3\} $, $\mathbf{p}_{u1}(i_1)>\mathbf{p}_{u1}(i_2)>\mathbf{p}_{u1}(i_3)$; $ \mathcal{I}_{u2}=\{i_4,i_5\} $, $\mathbf{p}_{u2}(i_4)>\mathbf{p}_{u2}(i_5)$. 
    Since $u_1$ and $u_2$ like similar items, such as $i_1$ and $i_4$, we give their distances large weights such that $D(u_1,u_2)$ is small. 
    In subfigure (b), since $u_3$ and $u_4$ like dissimilar items, such as $i_3$ and $i_5$, 
    we give their distances large weights such that $D(u_3,u_4)$ is large.}
    \label{fig:phy}
      \vspace{-1.5em}
 \end{figure*}
\section{Related Work}
COS and PCC are two most classic and widely-used user similarity measures. Afterwards, numerous variants of COS and PCC have been proposed, such as Adjusted Cosine ~\cite{sarwar2001item}, Constrained PCC~\cite{shardanand1995social}, Weighted PCC~\cite{ma2007effective}, Sigmoid PCC~\cite{jamali2009trustwalker}, etc. However, they are not motivated for handling the ``co-rated item" issue and still suffers from the problem.
Apart from them, some other measures, such as Jaccard~\cite{koutrika2009flexrecs}, MSD~\cite{shardanand1995social}, JMSD~\cite{bobadilla2010new}, URP~\cite{liu2014new}, NHSM~\cite{liu2014new}, PIP~\cite{ahn2008new} and BS~\cite{guo2013novel} also suffers from the ``co-rated item'' issue~\cite{al2017collaborative}. Although some of them, such as Jaccard and URP consider all ratings of each user, they are inefficient in use of the rating information. For example, Jaccard only uses the number of items while omits the specific rating values; URP only uses the mean and variance of rating values. Moreover, they all fail when there are no co-rated items at all.

The most relevant works to us are BCF~\cite{patra2015new} and HUSM~\cite{wang2017hybrid}, which predict user similarity by utilizing all ratings of each user and also item similarities. Although BCF and HUSM can partially solve the ``co-rated item'' issue and show good performance on sparse data, they also have drawbacks and sometimes give counter-intuitive results, as illustrated in section \ref{model} and section \ref{casestudy} respectively. For your convenience, please refer to the supplementary material for the definitions of these previous methods.


\section{The Proposed Similarity Measure}
\label{model}
\noindent \textbf{Problem definition.} Let $ \mathcal{U} $ be a set of $m$ users, and $ \mathcal{I} $ a set of $n$ items. 
The user-item interaction matrix is denoted by $ \mathbf{R} \in \mathbb{R}^{m\times n}$ with $\mathbf{R}(u,i) \geq 0$ the rating of user $u$ given to item $i$.
For user $u \in \mathcal{U}$, his rated items are $ \mathcal{I}_u \subset \mathcal{I}$.
Usually $ \mathbf{R}$ is a partially observed matrix and highly sparse. $d(i,j)\geq 0$ denotes the distance between item $i$ and $j$, which evaluates the similarity between them. The smaller $d(i,j)$ the more similar $i$ and $j$. We can derive $d(i,j)$ from ratings on items~\cite{patra2015new,wang2017hybrid} or content information~\cite{yao2018judging}, such as item tags, comments, etc. In this paper, we assume $d(i,j)$ are given. How to construct high-quality item distances is beyond the scope of this paper. We are interested in computing the distance between any pair of users in $ \mathcal{U}$ given $\mathbf{R}$ and $d$. Then user similarity can be easily derived from user distance, since they are negatively correlated.

\noindent \textbf{User preference representation.}
Let $\Sigma_k=\{\mathbf{p}\in[0,1]^k \;|\; \braket{\mathbf{p},\mathds{1}}=1\}$ denotes a $(k-1)$-dimensional simplex. 
We model a user $u$'s preference as a probabilistic distribution $\mathbf{p}_u\in \Sigma_{|\mathcal{I}_u |}$ on $ \mathcal{I}_u$, where $\mathbf{p}_u(i)$ represents how much the user $u$ likes item $i$. The larger $\mathbf{p}_u(i)$ the more $u$ likes $i$. In practice, the ground truth of $\mathbf{p}_u$ is unobserved and we estimate it by normalizing $u$'s ratings on $\mathcal{I}_u$, i.e. $\mathbf{p}_u(i) \approx \frac{\mathbf{R}(u,i)}{\sum_{j \in \mathcal{I}_u}\mathbf{R}(u,j)}$ for $i \in \mathcal{I}_u$.


\noindent \textbf{User preference distance.} 
Ratings and rated items both reflect the difference between a pair of users. Hence, we consider these two different types of information while modeling user distance.
Specifically, we model the distance between user $u$ and $v$, denoted by $D(u,v)$, as the weighted average of the distances among their rated items, i.e. 
\begin{align}\label{eq:1}
\sum_{i \in \mathcal{I}_u}\sum_{j \in \mathcal{I}_v}\mathbf{W}_{u,v}(i,j)d(i,j),
\end{align}
where $\mathbf{W}_{u,v}(i,j)\geq0$ is the weight of $d(i,j)$ and $\sum_{i \in \mathcal{I}_u}\sum_{j \in \mathcal{I}_v}\mathbf{W}_{u,v}(i,j)\\=1$. The weights $\mathbf{W}_{u,v}$ embody the difference among the ratings of $u$ and $v$ such that {$D(u,v)$ is small if $u$ and $v$ like (give high ratings to) similar items.}
Our strategy to construct such a $\mathbf{W}_{u,v}$ is: If $u$ and $v$ like similar items, the $\mathbf{W}_{u,v}(i,j)$ of similar items, whose $d(i,j)$ are small, should be large such that the resulting user distance is small; If $u$ and $v$ like dissimilar items, the $\mathbf{W}_{u,v}(i,j)$ of dissimilar items, whose $d(i,j)$ are large, should be large such that the resulting user distance is large. 
Figure \ref{fig:phy} briefly illustrates this strategy.
To summarize, we find the strategy just makes the mass of $\mathbf{W}_{u,v}$ concentrate on the item pairs that are `liked' by both users, as can be noticed in Figure \ref{fig:phy}a and \ref{fig:phy}b. 
Hence, we implement this strategy by making the marginal distribution of $\mathbf{W}_{u,v}$ follow $\mathbf{p}_u$ and $\mathbf{p}_v$ respectively, i.e. $\mathbf{W}_{u,v} \in U(\mathbf{p}_u,\mathbf{p}_v) $, where
\begin{align}\label{eq:2}
    U(\mathbf{p}_u,\mathbf{p}_v) :=& \left \{ \mathbf{W}_{u,v}\in[0,1]^{|\mathcal{I}_u|\times |\mathcal{I}_v|} \;|\;  \mathbf{W}_{u,v}\mathds{1}=\mathbf{p}_u, \mathbf{W}_{u,v}^T \mathds{1} =\mathbf{p}_v \right\}.
\end{align}
However, there are many $\mathbf{W}_{u,v}$ that satisfy $\mathbf{W}_{u,v} \in U(\mathbf{p}_u,\mathbf{p}_v) $, which results in the user distance value indeterminate. Therefore, we define the user distance as the smallest one among all possible values:
\begin{align}\label{eq:3}
    D(u,v) :=\min_{\mathbf{W}_{u,v}\in U(\mathbf{p}_u,\mathbf{p}_v)}  \sum_{i \in \mathcal{I}_u}\sum_{j \in \mathcal{I}_v}\mathbf{W}_{u,v}(i,j)d(i,j),
\end{align}
A benefit of taking minimum in formula (\ref{eq:3}) is that the resulting distance is a metric as long as $d$ is a metric~\cite{rubner1998metric}. 
\begin{table*}
\small
\begin{subtable}[t]{0.48\textwidth}
\begin{tabular}{|c|c|c|c|c|c|}
\hline
     & Iron Man & Bat Man & Spider Man & Titanic & Casablanca \\ \hline
$u_1$ & 5        & ---     & 2          & 3       & ---        \\ \hline
$u_2$ & ---      & 5       & 2          & 3       & ---        \\ \hline
$u_3$ & ---      & ---     & 2          & 3       & 5          \\ \hline
$u_4$ & 5        & 5       & 5          & ---     & ---        \\ \hline
$u_5$ & ---      & ---     & ---        & 5       & ---        \\ \hline
$u_6$ & ---      & ---     & ---        & ---     & 5          \\ \hline
\end{tabular}
\caption{Users' ratings on five movies.}
\label{tab:user}
\end{subtable}
\begin{subtable}[t]{0.48\textwidth}
\begin{tabular}{|c|c|c|c|c|c|}
\hline
           &  \makecell{Iron\\Man} & \makecell{Bat\\Man} & \makecell{Spider\\Man} & Titanic & Casablanca \\ \hline
Iron Man   & 1        & 0.8     & 0.8        & 0.3     & 0.3        \\ \hline
Bat Man    & 0.8      & 1       & 0.8        & 0.3     & 0.3        \\ \hline
Spider Man & 0.8      & 0.8     & 1          & 0.3     & 0.3        \\ \hline
Titanic    & 0.3      & 0.3     & 0.3        & 1       & 0.8        \\ \hline
Casablanca & 0.3      & 0.3     & 0.3        & 0.8     & 1          \\ \hline
\end{tabular}
\caption{The similarity matrix among movies. For PMD, we derive $d(i,j)$ by $d(i,j):=arccos(similarity(i,j))$.}
\label{tab:item_sim}
\end{subtable}
\hspace{\fill}
\centering
\begin{subtable}[t]{1\textwidth}\centering
\begin{tabular}{|c|c|c|c|c|c|c|c|c|c|c|c|c|c|}
\hline
Case&\makecell{Similarity\\between} 
                    & COS & PCC & $1-$MSD & Jaccard &URP & JMSD & NHSM & BCF & N-BCF & HUSM &N-HUSM& $1-$PMD \\ \hline
\multirow{2}{5em}{(1) Co-rated items exist }&$u_1$ \& $u_2$     & 1    & 1    & 1 & 0.5      &  0.5&  0.5&  0.0769 & 3.277* &0.809*&0.197*&0.0218*&0.892*  \\ \cline{2-14} 
&$u_2$ \& $u_3$     & 1    & 1    & 1 & 0.5      &  0.5&  0.5&  0.0769 & 2.444* &0.716*&0.157*&0.0175*&0.633* \\ \hline 
\hline
\multirow{2}{5em}{(2) No co-\\rated items}&$u_4$ \& $u_5$     & --- &  --- & --- &  0 & 0.5 & --- & --- & 0.9 &0.3*&0.0552&0.0184*& 0.3* \\ \cline{2-14} 
&$u_5$ \& $u_6$     & --- &  --- & --- &  0 & 0.5 & --- & --- & 0.8 &0.8*&0.0491&0.0491*& 0.8* \\ \hline
\end{tabular}
\caption{User similarity computed by various methods. Since MSD and PMD are distance measures, we convert them to similarity values by $similarity:=1-distance$ for easy comparison. Desirable values are followed by a `*'.}
\label{tab:val}
\end{subtable}
\hspace{\fill}
\vspace{-1em}
\caption{Toy examples of user similarity computed by various methods.}
\label{tab:case}
\vspace{-2.7em}
\end{table*}
The optimization in formula (\ref{eq:3}) is in coincidence with a special case of the earth mover’s distance metric (EMD)~\cite{monge1781memoire,rubner1998metric,wolsey2014integer}, a well studied
transportation problem for which specialized solvers have
been developed~\cite{ling2007efficient,pele2009fast}. 
To highlight this connection we refer to our proposed user distance as the Preference Mover’s Distance (PMD).

\noindent \textbf{Remark 1:} Comparing with BCF and HUSM.
BCF and HUSM also have a similar basic form of $\sum_{i \in \mathcal{I}_u}\sum_{j \in \mathcal{I}_v}Sim_{rating}(r_{ui},r_{vj})\cdot \\Sim_{item}(i,j)$, where $Sim_{rating}(r_{ui},r_{vj})$ evaluates rating similarity and $Sim_{item}(i,j)$ represents item similarity. 
Nevertheless, (1) They don't normalize the number of terms in the summation. Therefore, the resulting user similarity can grow linearly w.r.t. the number of terms (item pairs) in the sum. However, the number of item pairs do not necessarily imply user similarity. Thus, the resulting similarity values could be misleading. In contrast, $\mathbf{W}_{u,v}$ serves as a natural normalization for PMD.
(2) Even if we consider their normalized versions, BCF and HUSM derive each $Sim_{rating}(r_{ui},r_{vj})$ independently and heuristically, while we derive $\mathbf{W}_{u,v}$ through optimization. As a result, PMD satisfies triangle inequality, while BCF and HUSM do not have equivalent properties. 
Triangle inequality limits each distance value to a reasonable range which is defined by other distance values. 
Consequently, PMD can guarantee that if user A and B are both similar to C, then A and B are also similar to each other.
Thus, we can weigh the neighbors more reasonably according to the distance values given by PMD.


\section{Experiments}
In this section, we first use several case studies to give the reader an insight into the superiority of PMD over other measures, then we compare the recommendation accuracy of these measures on two real world data sets.

\noindent \textbf{Comparison Measures.} COS, PCC and MSD are three classic user similarity measures. 
Jaccard, URP, JMSD, NHSM, BCF, HUSM are five representatives that try to use all ratings of each user to evaluate user similarity. 
To study how normalization influence the performance of BCF and HUSM, we also consider their normalized versions which normalize the number of item pairs: \\$\frac{\sum_{i \in \mathcal{I}_u}\sum_{j \in \mathcal{I}_v}Sim_{rating}(r_{ui},r_{vj})\cdot \\Sim_{item}(i,j)}{|\mathcal{I}_u|\times|\mathcal{I}_v|}$. The resulting similarity measures are named as \textit{N-BCF} and \textit{N-HUSM} respectively.



\subsection{Case Study}
\label{casestudy}
We illustrate by examples the differences among the similarity values computed by various methods under the two cases: (1) there exist co-rated items and (2) there are no co-rated items. 
In Table \ref{tab:user}, the ratings of six users on five movies are presented.
User $u_1$, $u_2$ and $u_3$ are used to simulate case (1) while user $u_3$, $u_4$ and $u_5$ to simulate case (2).
The five movies consist of three sci-fi movies and two romantic movies. 
As BCF, HUSM and PMD require item similarity/distance for computation, we assume we know the ground truth of movie similarities, as shown in \ref{tab:item_sim}.
The similarity values  computed by various methods are shown in \ref{tab:val}.



In Case (1), although $u_1$, $u_2$ and $u_3$ all give exactly the same ratings to their co-rated items, $u_1$ and $u_2$'s high ratings are both on sci-fi movies while $u_3$'s high ratings on romantic movies. This implies that  $u_1$ and $u_2$ like sci-fi more than romance while $u_3$ loves romance more. Therefore, $u_1$ and $u_2$ should be more similar than $u_2$ and $u_3$. 
However, COS, PCC, MSD, Jaccard, URP, JMSD and NHSM all give $Sim(u_1,u_2)=Sim(u_2,u_3)$, which is counter-intuitive.
The reason is that COS, PCC, MSD only consider ratings on co-rated movies; Jaccard omits the exact rating values; and URP only considers mean and variance of rating values. 
JMSD and NHSM also fail in this case because their ability of using all ratings is granted by Jaccard or URP. 
The only successful measures in this case are BCF, N-BCF, HUSM, N-HUSM and our PMD, because they make full use of all rating information.

In Case (2), $u_3$, $u_4$ and $u_5$ have no co-rated items at all. However, $u_3$ and $u_4$'s high ratings are both on sci-fi movies while $u_5$'s high ratings on romantic movies. Like Case (1), $u_3$ and $u_4$ should be more similar than $u_4$ and $u_5$. However, COS, PCC, MSD, JMSD and NHSM all can't work in this case because they must use the ratings on co-rated items.
Jaccard gives 0 as long as there are no co-rated items.
URP again gives identical similarity values for the same reason discussed in Case (1).
Although BCF and HUSM utilize all rating information, they give a misleading result of $Sim(u_4,u_5)>Sim(u_5,u_6)$ while the desired one should be $Sim(u_4,u_5)<Sim(u_5,u_6)$.
This is because there are more item pairs between $u_4$ and $u_5$ than $u_5$ and $u_6$ and they didn't normalize the number of item pairs. By contrast, N-BCF and N-HUSM works well because of normalization. 
Our PMD again performs well even in the absence of co-rated items and is not misled by the number of item pairs.

Although N-BCF, N-HUSM and PMD can all judge user similarity correctly in Case (1) and (2), PMD generates more reasonable distance values as discussed in Remark 1. In the next section, we conduct experiments to further justify this point.
\begin{figure*}[t]
    \centering
    \begin{subfigure}[b]{0.25\textwidth}
        \includegraphics[width=\textwidth]{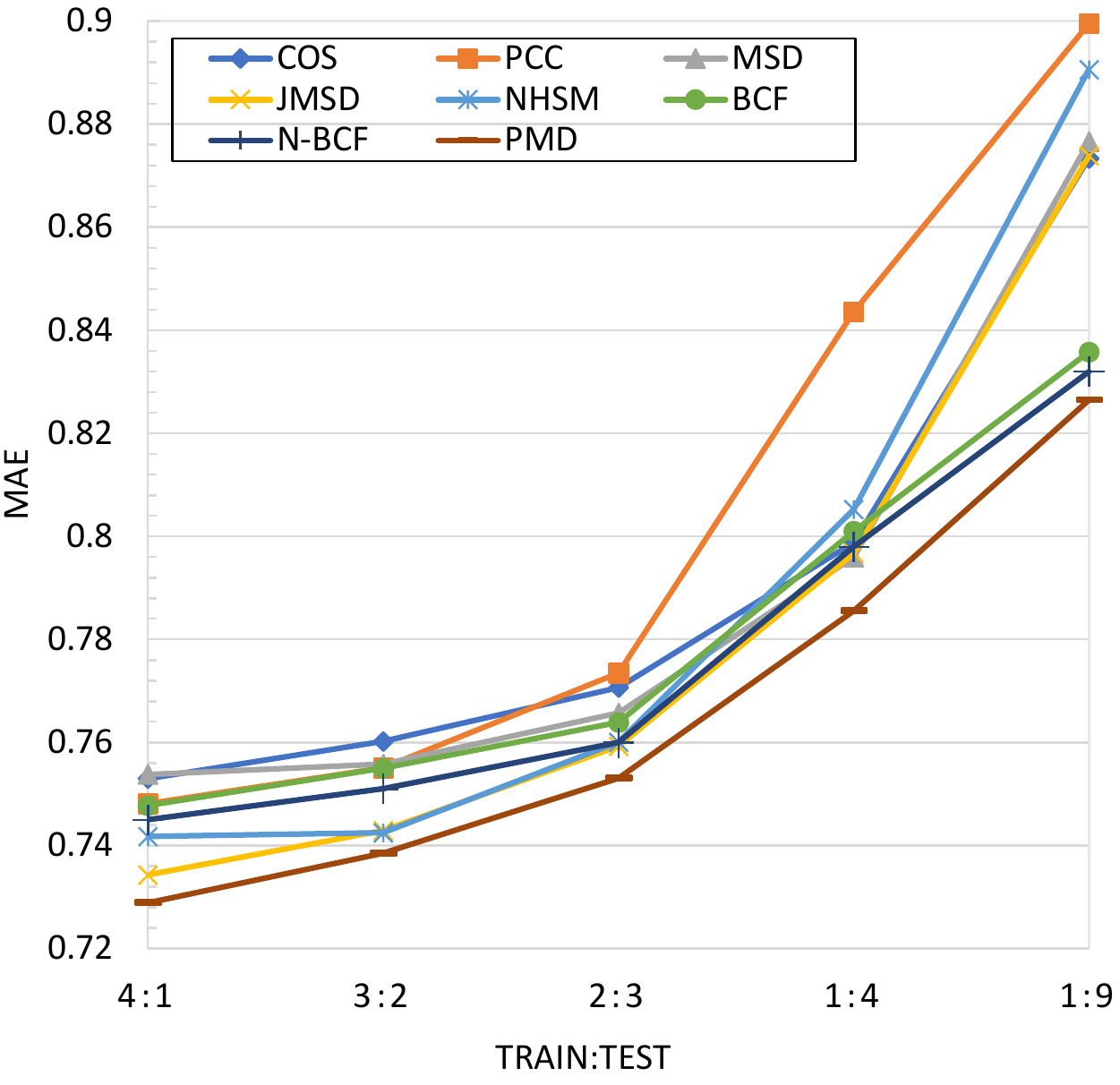}
        \caption{train:test from 4:1 to 1:9; K=40.}
        \label{fig:100k_tr}
    \end{subfigure}
    \quad
    \begin{subfigure}[b]{0.25\textwidth}
        \includegraphics[width=\textwidth]{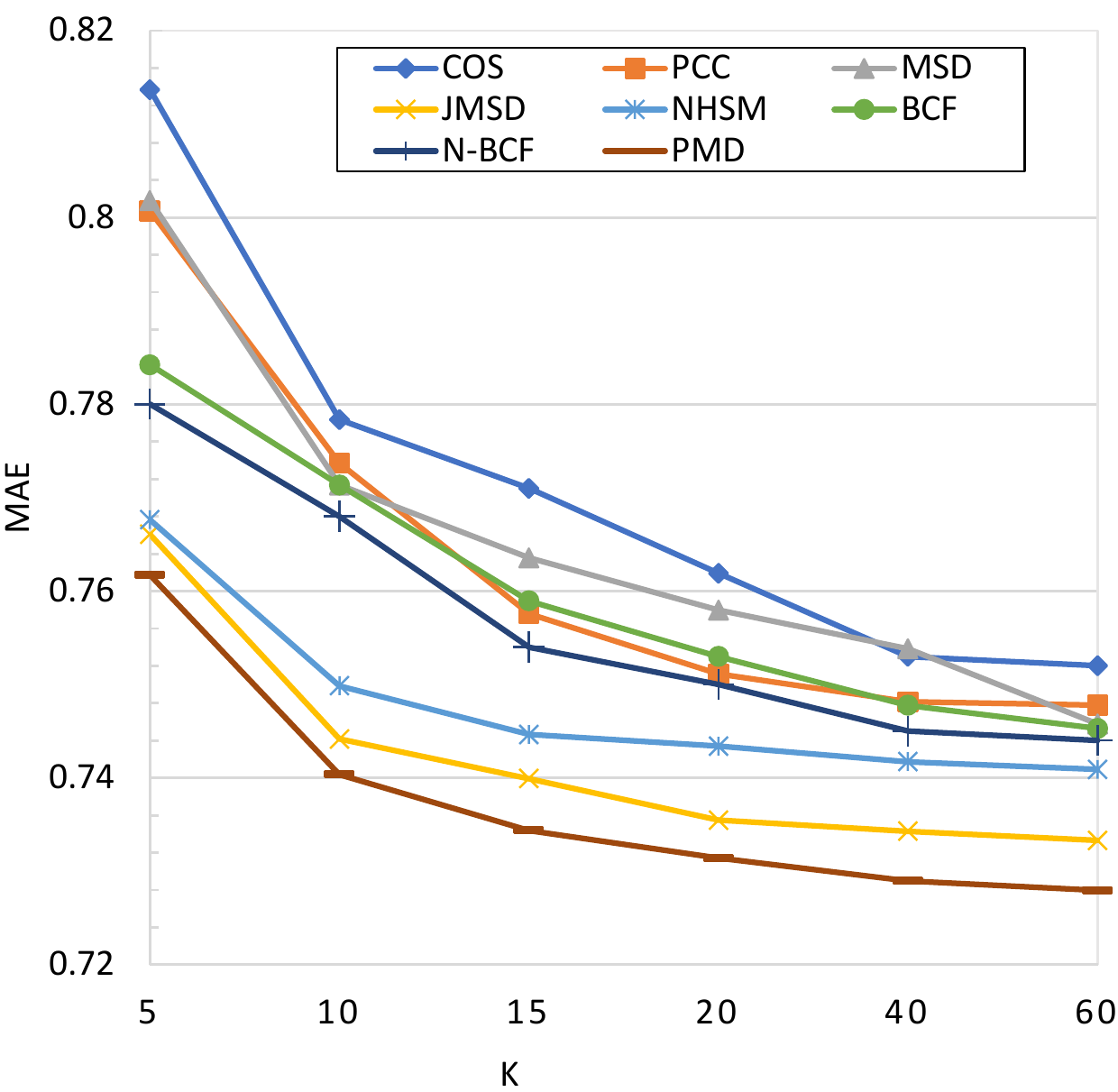}
        \caption{train:test=4:1; K from 5 to 60.}
        \label{fig:100k_4_1}
    \end{subfigure}
    \quad
    \begin{subfigure}[b]{0.25\textwidth}
        \includegraphics[width=\textwidth]{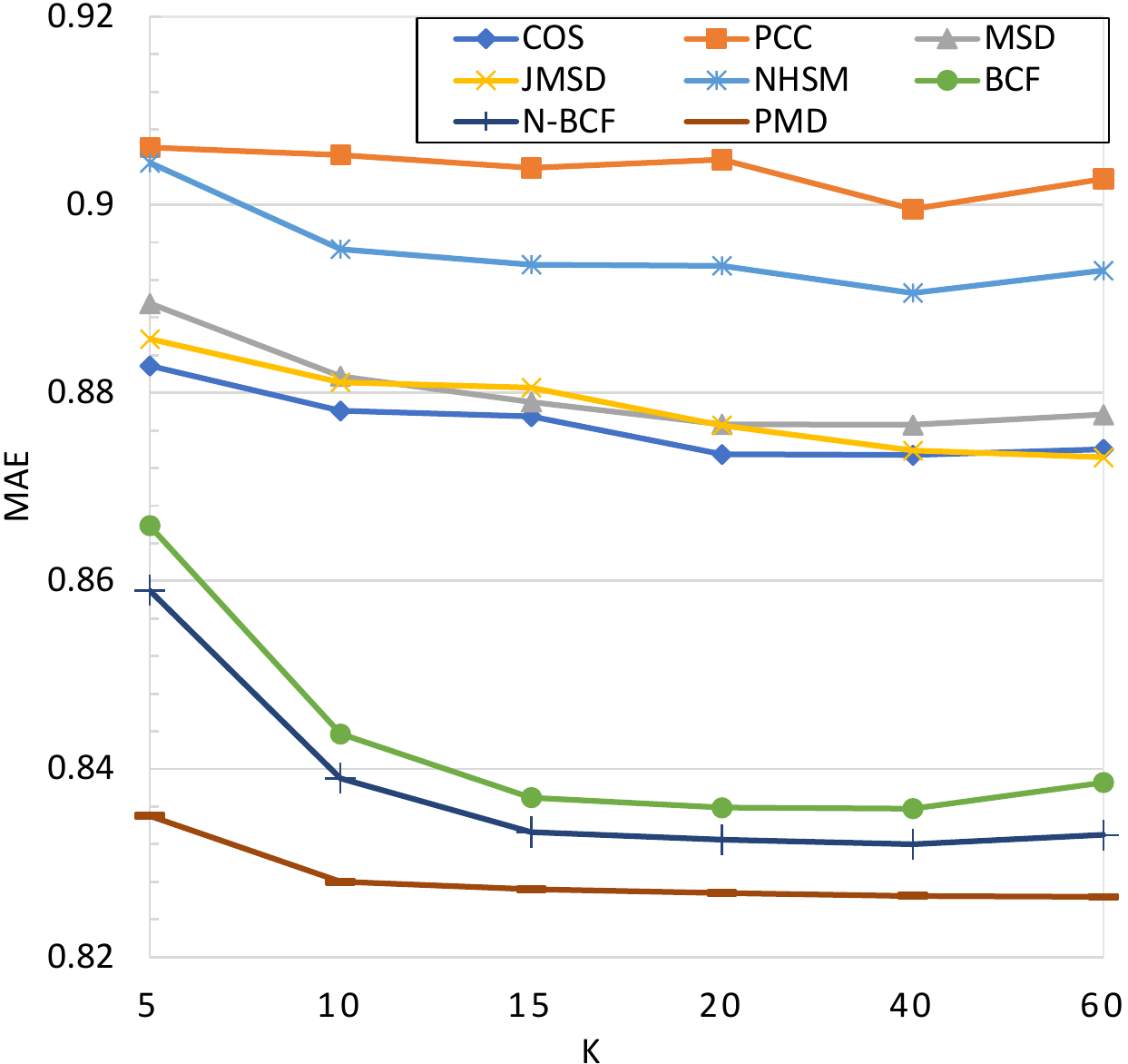}
        \caption{train:test=1:9; K from 5 to 60.}
        \label{fig:100k_1_9}
    \end{subfigure}
    \quad
    \begin{subfigure}[b]{0.25\textwidth}
        \includegraphics[width=\textwidth]{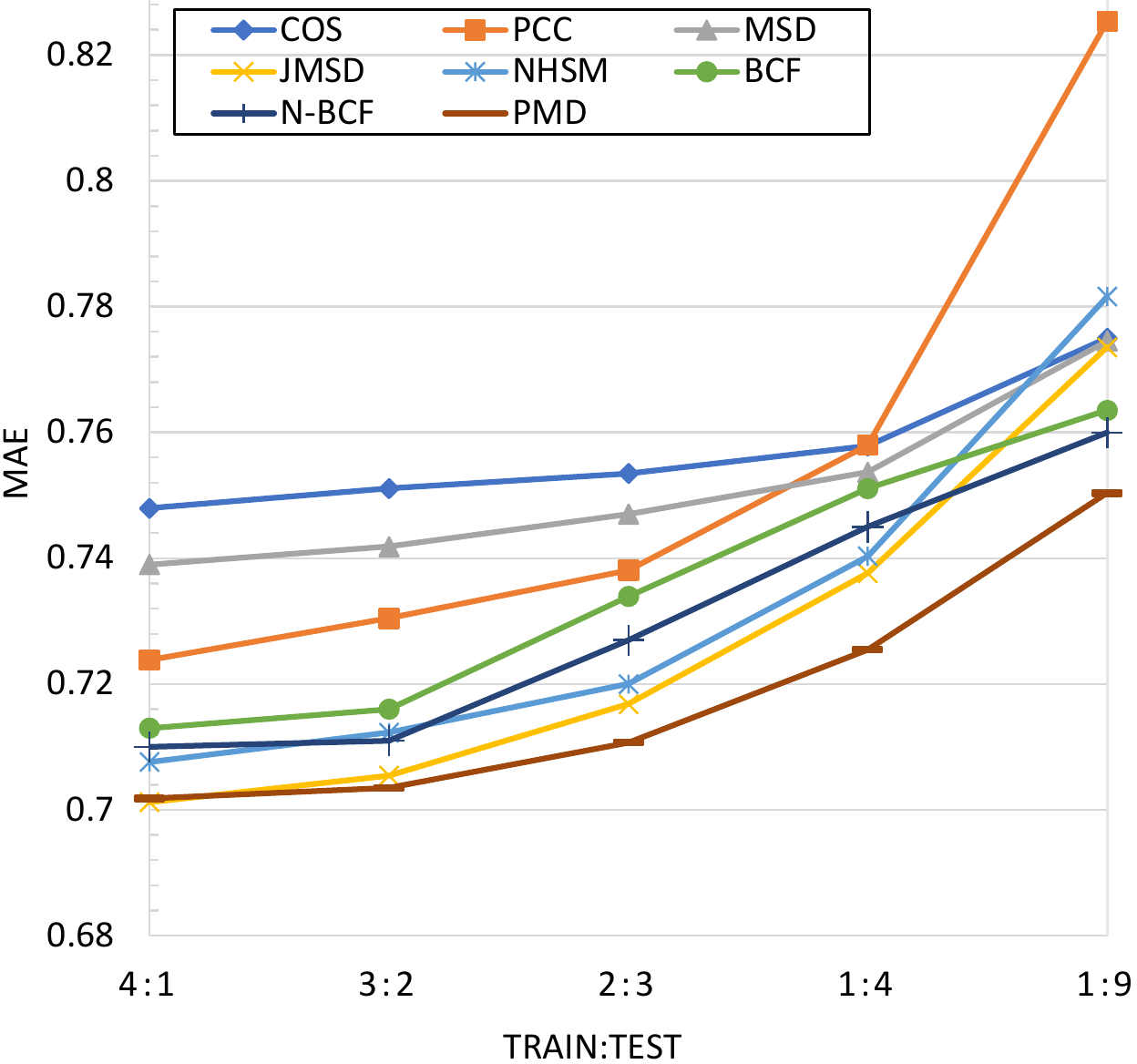}
        \caption{train:test from 4:1 to 1:9; K=40.}
        \label{fig:1m_tr}
    \end{subfigure}
      \quad
    \begin{subfigure}[b]{0.25\textwidth}
        \includegraphics[width=\textwidth]{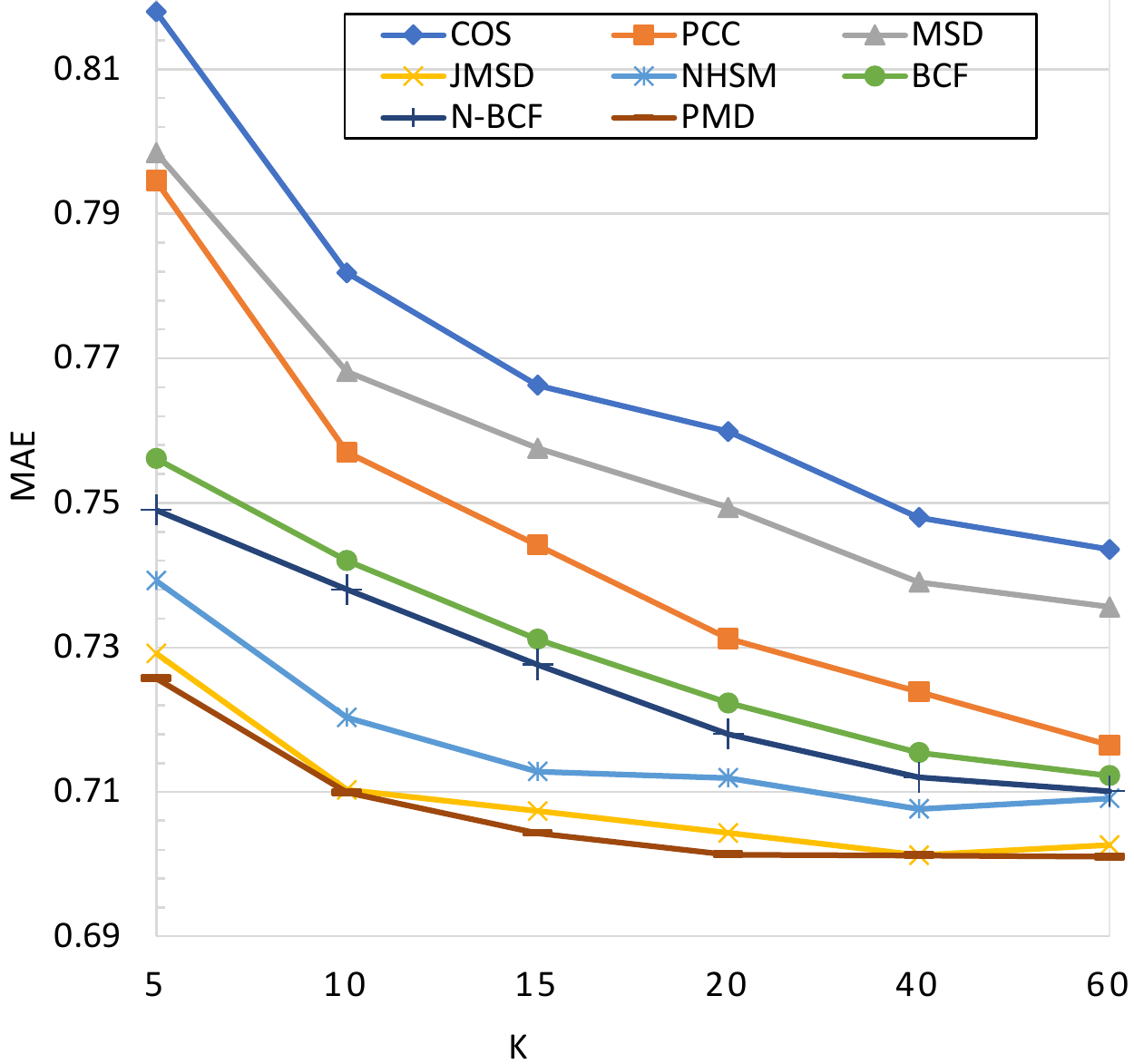}
        \caption{train:test=4:1; K from 4 to 60.}
        \label{fig:1m_4_1}
    \end{subfigure}
    \quad
    \begin{subfigure}[b]{0.25\textwidth}
        \includegraphics[width=\textwidth]{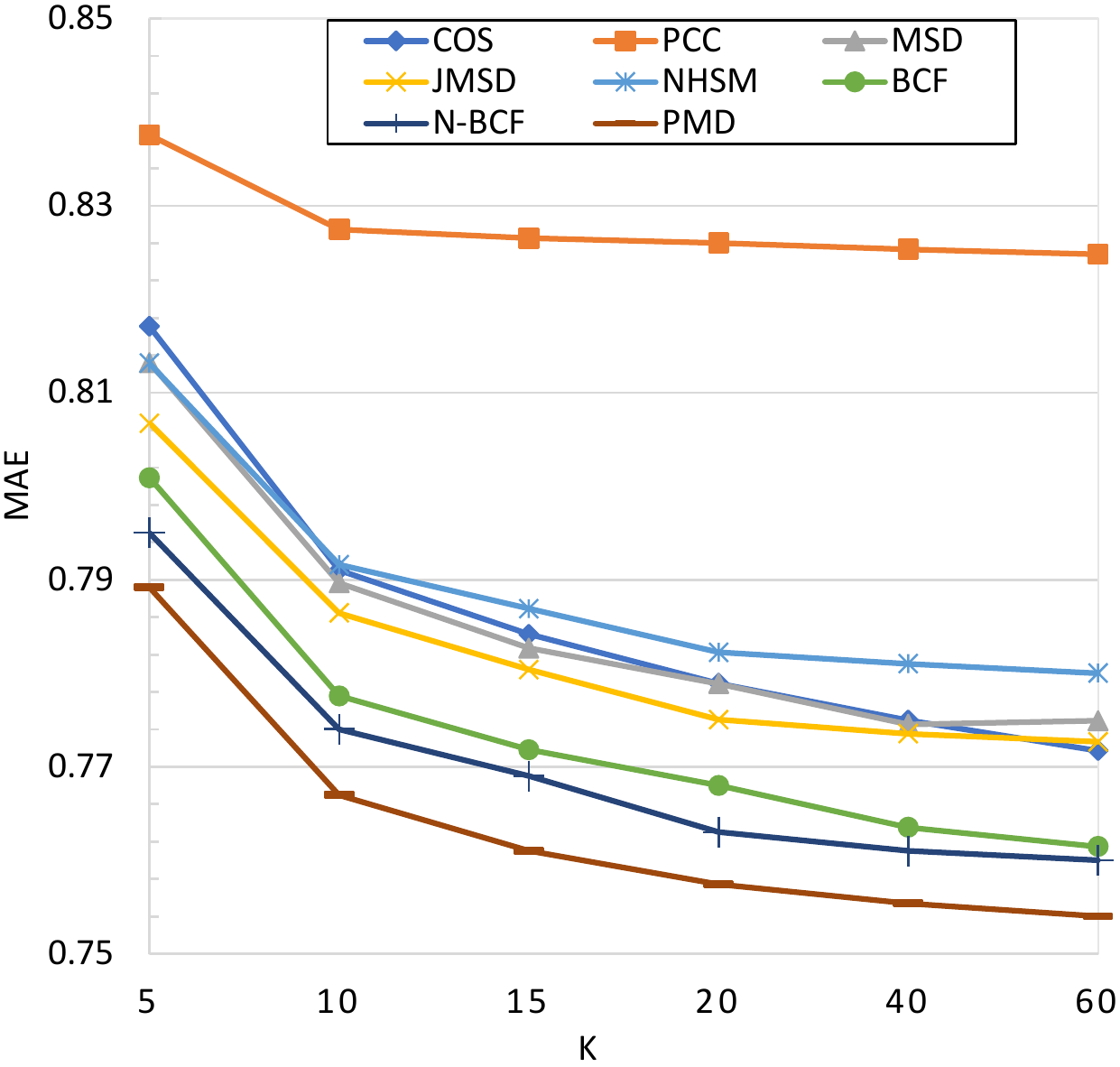}
        \caption{train:test=1:9; K from 4 to 60.}
        \label{fig:1m_1_9}
    \end{subfigure}
    \caption{MAE of comparative approaches on MovieLens- 100K (a-c) and MovieLens-1M (d-f). Experimental settings corresponding to each subfigure are summarized in their subtitles in a format of "train:test ratio; K."}
    \label{fig:result}
\vspace{-1em}
\end{figure*}
\subsection{Experiments on Real World Data Set.}
\label{data}
We evaluate the recommendation performance of various similarity measures by cross validation on two well-known datasets: MovieLens-100k and MovieLens-1M\footnote{The two datasets are available at \url{https://grouplens.org/datasets/movielens/}}. 
We perform each experiment five times and take the average test performance as the final results.
We apply the K-NN approach to select a group of similar users whose ranking are in the top K according to user similarity/distance. The ratings of selected similar users are aggregated to predict items’ ratings by a mean-centring approach~\cite{desrosiers2011comprehensive}.
We use mean absolute error (MAE) to evaluate recommendation performance.
Lower MAE indicates better accuracy. While our experiments
use memory-based CF, we emphasize that user distance computation is equally relevant to model-based methods, including those based on matrix factorization such as~\shortcite{ma2011recommender}.


\noindent \textbf{Item similarity}
We construct item similarity matrix by 
Tag-genomes\footnote{\url{https://grouplens.org/datasets/movielens/tag-genome/}} ~\cite{vig2012tag} of movielens.org. For the sake of fairness, the same item similarity matrix is used in PMD, BCF, N-BCF, HUSM and N-HUSM\footnote{BCF and HUSM originally compute item similarity by Bhattacharyya coefficient or KL-divergence of ratings, but we found the resulting performance is inferior to that by using tag-genomes, possibly because tag-genomes can better describe a movie than the ratings.}.
We compute the cosine value of the tag-genomes of two movies as the similarity between them, which is a well-developed method to evaluate movie similarities~\cite{vig2012tag,yao2018judging}. 
For PMD, it needs to convert item similarity to item distance and we define the item distance as $Distance:=arccos(Similarity)$, which is a metric. 
There are of course other ways~\cite{yao2018judging} to construct item similarity matrix. We use tag-genome because it gives item similarity of high quality~\cite{yao2018judging}. How to construct high quality item similarities is beyond the scope of this paper.


\noindent \textbf{Experimental Protocol.}
To test the performance of PMD and its robustness to data sparsity, we vary the train to test ratio from 4:1 to 1:9, during which the training data become more and more sparse.
In addition, in order to study how parameter K affects the performance of various methods, we vary K from 5 to 60 with train:test ratio fixed to specific values.


\noindent \textbf{Results.}
Experimental results are summarized in Figure \ref{fig:result}
\footnote{For the neat of presentation, we only present the performance of 7 most competitive and representative baselines.}.

Figure \ref{fig:100k_tr} and \ref{fig:1m_tr} shows how different levels of data sparsity affects the performance of various methods when K is fixed to 40. 
PMD outperforms all competitive methods on all train to test ratios. 
In the two figures, the performance of all measures degrade as the train to test ratio deceases, because training data is becoming more and more sparse.
Specifically, the performance gaps between our method and those of COS, PCC, MSD, JMSD and NHSM become larger and larger as the training data grows more and more sparse. This is probably because when data goes sparse, co-rated items hardly exist for most user pairs, making the resulting user similarity misleading or uncomputable for these methods.
By contrast, PMD can compute similarity properly no matter co-rated items exist or not. 
When train to test ratio is 1:9, PMD can significantly advance these measures, which is also notable in \ref{fig:100k_1_9} and \ref{fig:1m_1_9}.
BCF and N-BCF shows a similar trend as PMD and beats all previous methods when data is highly sparse (e.g. train:test=1:9), probably because they also make full use of all ratings of each user. N-BCF performs slightly better than BCF possibly because of normalization.
However, PMD always performs better than BCF and N-BCF, possibly because PMD has a natural normalization strategy and can weigh the neighbors more reasonably as discussed in Remark 1.

Figure \ref{fig:100k_4_1} and \ref{fig:1m_4_1} show how K affects the performance when train to test ratio is fixed to 4:1. As the number of neighbors increases from 5 to 60, the performance of all methods improves. This may be because more information are incorporated and noise is averaged out.
PMD outperforms others consistently when K varies. 
This again shows PMD can give more accurate user similarity than other methods.

Similarly, figure \ref{fig:100k_1_9} and \ref{fig:1m_1_9} show how K affects the performance when the train to test ratio is 1:9, which means the training data is even more sparse. Our method shows significant advantage over the competitive baselines. In \ref{fig:100k_1_9}, the performance of COS, PCC, MSD, JMSD and NHSM are nearly constant as K varies and stay at a low accuracy level. This may be because many user pairs have no co-rated items and consequently the corresponding similarity is uncomputable when training data is highly sparse. Thus, for many users, the number of neighbors with an effective similarity value is often less than the parameter K, resulting the performance unchanged when K increases.

\bibliographystyle{ACM-Reference-Format}
\bibliography{sample-bibliography}

\end{document}